\documentclass[atoms,article,submit,moreauthors,pdftex,12pt,a4paper]{mdpi} 
\lastpage{x}
\doinum{10.3390/------}
\pubvolume{xx}
\pubyear{2015}
\history{Received: xx / Accepted: xx / Published: xx}
\pdfoutput=1

\usepackage[normalem]{ulem}
\usepackage[usenames,dvipsnames]{xcolor}

\renewcommand{\aa}{\hat{a}}
\newcommand{\adagger}{\hat{a}^\dagger}

\newcommand{\ppsi}{\hat{\psi}}
\newcommand{\psidagger}{\hat{\psi}^\dagger}
\newcommand{\n}{\hat{n}}

\usepackage[caption=false]{subfig}

   \newcommand{\ket}[1]{\left|#1\right>} \newcommand{\bra}[1]{\left<#1\right|}
   \newcommand{\braket}[2] {\left<#1|#2\right>} 
   
\newcommand{\qav}[1]{\langle #1\rangle}

\newcommand{\ba}{\begin{align}}
\newcommand{\ea}{\end{align}}

\newcommand{\eea}{\end{eqnarray}}

\newcommand{\qavg}[1]{\ensuremath{\langle #1 \rangle}}
\newcommand{\parlr}[1]{\ensuremath{\left(#1\right)}}
\newcommand{\sqlr}[1] {\ensuremath{\left[ #1 \right]}}

\newcommand{\ha}{\ensuremath{\hat{a}}}
\newcommand{\hb}{\ensuremath{\hat{b}}}
\newcommand{\had}{\ensuremath{\hat{a}^{\dagger}}}
\newcommand{\hbd}{\ensuremath{\hat{b}^{\dagger}}}

\newcommand{\hpsi}{\ensuremath{\hat{\Psi}}}
\newcommand{\hpsid}{\ensuremath{\hat{\Psi}^{\dagger}}}

\newcommand{\hd}{\ensuremath{\hat{d}}}
\newcommand{\hdd}{\ensuremath{\hat{d}^{\dagger}}}
\newcommand{\hn}{\ensuremath{\hat{n}}}

\makeatletter

\newcommand{\Rmnum}[1]{\expandafter\@slowromancap\romannumeral #1@}
\makeatother

\Title{An optomechanical elevator: Transport of a Bloch oscillating Bose-Einstein condensate up and down an optical lattice by cavity sideband amplification and cooling.}

\Author{B. Prasanna Venkatesh $^{1,4,5}$*, D.H.J. O'Dell $^{2}$ and J. Goldwin $^{3}$}

\address{%
$^{1}$ Asia Pacific Center for Theoretical Physics, San 31, Hyoja-dong, Nam-gu, Pohang, Gyeongbuk 790-784, Korea\\
$^{2}$ Department of Physics and Astronomy, McMaster University, 1280 Main Street West, Hamilton, Ontario L8S 4M1, Canada\\
$^{3}$ Midlands Ultracold Atom Research Centre, School of Physics and Astronomy, University of Birmingham, Edgbaston, Birmingham B15 2TT, United Kingdom\\
$^{4}$ Institute for Quantum Optics and Quantum Information of the Austrian Academy of Sciences, Technikerstra\ss e 21a, Innsbruck 6020, Austria\\
$^{5}$ Institute for Theoretical Physics, University of Innsbruck, A-6020 Innsbruck, Austria}
\corres{Prasanna.Venkatesh@uibk.ac.at}

\abstract{\nolinenumbers{{In this paper we give a new description, in terms of optomechanics, of previous work on the problem of an atomic Bose--Einstein condensate interacting with the optical lattice inside a laser-pumped optical cavity and subject to a bias force, such as gravity. \mbox{An atomic} wave packet in a tilted lattice undergoes Bloch oscillations; in a high-finesse optical cavity the backaction of the atoms on the light leads to a time-dependent modulation of the intracavity lattice depth at the Bloch frequency which can in turn transport the atoms up or down the lattice.} In the optomechanical picture, the transport dynamics can be interpreted as a manifestation of dynamical backaction-induced sideband damping/amplification of the Bloch oscillator. \mbox{Depending on} the sign of the pump-cavity detuning, atoms are transported either with or against the bias force accompanied by an up- or down-conversion of the frequency of the pump laser light. \mbox{We also} evaluate the prospects for using the optomechanical Bloch oscillator to make continuous measurements of forces by reading out the Bloch frequency. In this context, we establish the significant result that the optical spring effect is absent and the Bloch frequency is not modified by the backaction.}}


\begin{document}
\nolinenumbers{

\section{Introduction}

In an optomechanical system, light couples to mechanical degrees of freedom via radiation pressure. The classic setup consists of an optical cavity made of two mirrors, one of which has a very low mass and is suspended from a spring or pendulum \cite{Kip07,Asp14}; see Figure \ref{fig:fig1}. When pumped by an external laser, which is quasi-resonant with one of its modes, a large amplitude optical field builds up in the cavity, and the mobile mirror can be displaced as a result. This radiation pressure-induced lengthening of a cavity was observed by Dorsel \emph{et al}. in 1983 \cite{Dor83} and can become important in high precision optical interferometers, like those built to detect gravitational \mbox{waves \cite{Braginsky+Manukin77,Braginsky01,Braginsky02,Corbitt06}}. \mbox{Furthermore, because} a displacement of the mirror shifts the mode frequency, and therefore its detuning from the pump laser, and therefore the amplitude of the cavity field, there is a natural feedback loop acting on the radiation pressure. This backaction can cool the mirror if there is a phase delay between the oscillations of the mirror and the light \cite{Braginsky+Manukin77}, as occurs in good cavities where the relaxation time due to light leaking out of the cavity becomes long enough to be comparable to the period of the mirror motion.
The tantalizing possibility of cooling a macroscopic object, such as a mirror, to its quantum mechanical ground state has spurred remarkable experimental progress on a variety of optomechanical systems over the last ten years \cite{AspelmeyerZeilinger06,arcizet06,Schliesser06,VahalaRokhsari,KarraiFavero07,Schliesser08,thompson08,HarrisJayich08,Painter09,Groblacher09,Park09,Schliesser09,Rocheleau10,Sankey2010,KippenbergRiviere,Chan11,LehnertRegal1}. These include mirrors attached to cantilevers \cite{KarraiFavero07,AspelmeyerZeilinger06}, mechanical oscillators in microwave and optical cavities \cite{LehnertRegal1}, ``membrane-in-the-middle'' cavities \cite{thompson08,HarrisJayich08,Sankey2010}, ultra-high-Q microtoroids (i.e.\ toroids made of silica fused to cylindrical pillars made of silicon) \cite{VahalaRokhsari}, microspheres \cite{Park09} and optomechanical crystals \cite{Painter09}. In the latter three experiments, the mechanical oscillator modes were elastic shape deformations of the device itself.

\begin{figure}[H]
\centering
\includegraphics[height=5cm]{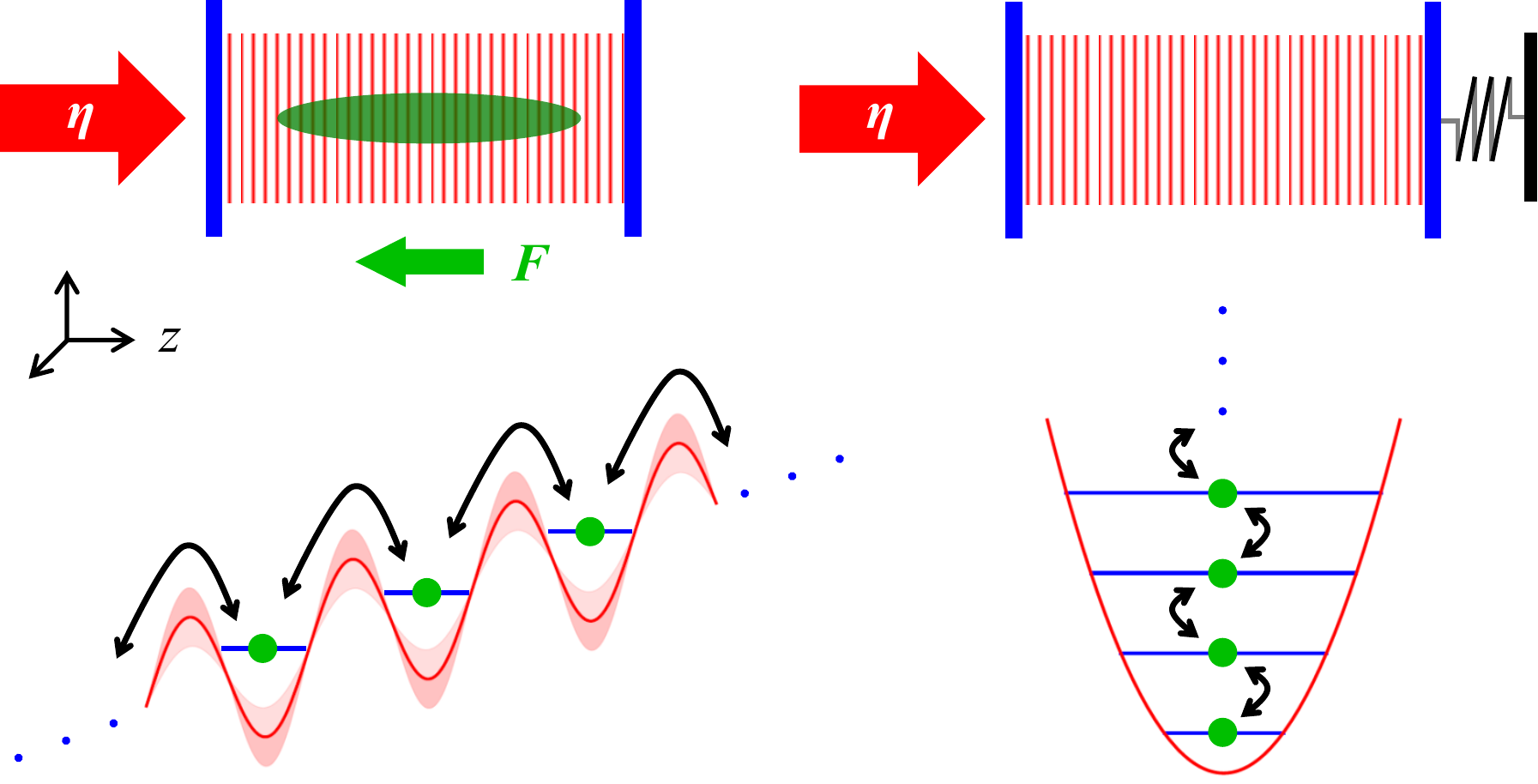}
\caption{(\textbf{a}) {An optical lattice} is created by pumping a Fabry--Perot cavity. The bias force $F$ tilts the lattice potential and causes a trapped Bose--Einstein condensate (green) to undergo Bloch oscillations. Atomic backaction leads to a time-modulation of the lattice amplitude, which in turn induces coherent directed transport of the condensate, provided the total detuning from resonance is not zero. \mbox{The transport} corresponds to atoms climbing or descending the ladder of Wannier--Stark states, where each state is separated by the energy $\hbar \omega_{B}$. (\textbf{b}) The archetype of cavity optomechanics is a cavity with one end mirror suspended on a spring. The motional states of the mirror correspond to the excitations of a harmonic oscillator. Cavity amplification or cooling of the mirror moves it up or down the ladder of oscillator states. Bloch oscillation dynamics in a cavity can be mapped onto standard cavity optomechanics, but with the harmonic oscillator ladder replaced by the Wannier--Stark ladder. } 
\label{fig:fig1}
\end{figure}

Another system in which optomechanics plays a central role consists of a gas of cold atoms trapped inside an optical cavity \cite{Ritsch13,Murch08,Brennecke08,Wolke2012}. In contrast to the experiments listed above, which all involved material oscillators (albeit, some of them weighing only nanograms \cite{AspelmeyerZeilinger06}) starting off well above their motional ground states, in the atomic gas experiments, the atoms are pre-cooled by a combination of laser cooling and forced evaporation, so that they start off at essentially zero temperature. This allows quantum effects to take centre stage. In the experiment by Murch \emph{et al}., the role of the mechanical oscillator was played by the centre-of-mass oscillations of atom clouds trapped in the wells of an intracavity optical lattice (standing wave of laser light in the cavity) \cite{Murch08}. The experiment by Brennecke \emph{et al}. was based on a Bose--Einstein condensate (BEC), which formed an effective two-state system: a lower state associated with the unperturbed BEC and an excited state given by a collective density wave excitation (phonon) created by the interaction of the BEC with the intracavity optical lattice \cite{Brennecke08}. When comparing the Hamiltonian of this system to that of the standard ``cavity-mirror-on-a-spring'' optomechanical system, one finds that it is the phonon that plays the role of the mechanical oscillator. Another interesting development in optomechanics is given by hybrid atom-membrane optomechanics \cite{Treutlein1,MeystreBariani}, where trapped atom clouds are coupled to a membrane using an optical lattice.

In this paper, we likewise consider a gaseous BEC trapped inside an optical cavity, which is pumped by a laser through one of the end mirrors. However, what distinguishes our system from the preceding ones is the inclusion of an extra force $F$ that acts on the atoms; this could be due to gravity or some other force, which is assumed to be constant over the length of the atom cloud. Rather than uniformly accelerating under the influence of $F$, in the presence of an optical lattice, the atoms undergo Bloch oscillations \cite{BenDahan96,BenDahan96-2,wilkinson96}. As shown by us in previous work \cite{Ven09}, this setup allows for a non-destructive measurement of $F$ if the backaction of the Bloch oscillating atoms on the cavity field is strong enough to imprint a periodic signal, which can be detected in the light leaking out of the cavity  (a related scheme has been independently proposed by Peden \emph{et al}. \cite{Ped09}).
{A number of experiments have already demonstrated how the light transmitted by a cavity can be used to track the motion of atoms trapped \mbox{inside \cite{Hood98,Hood00,Pinkse00},} and in particular, a theoretical analysis of the information stored in the frequency spectrum has been given \mbox{in \cite{Gangl00},} showing that atomic motion introduces sidebands on either side of the pump frequency. In our case, Bloch oscillations at angular frequency $\omega_{B}$ generate sidebands separated from the pump frequency by $\pm \omega_{B}$ (and harmonics thereof in the strong coupling regime). \mbox{Because the} Bloch frequency is proportional to the applied force $\omega_{B}= F d/\hbar$, where $d=\lambda/2$ is the lattice period, a detection of the spectrum of the transmitted light gives $F$ directly. }

Experiments using Bloch oscillating atoms to measure gravity are already well developed in optical lattices in {free space} (no cavity), where there is negligible backaction on the light by the \mbox{atoms \cite{Roati04,Ferrari06,Alb10,Poli11,Tar12}.} In these experiments, the momentum of the atoms is deduced from a destructive time-of-flight measurement requiring the lattice to be switched off after a variable hold time and the atom cloud imaged following ballistic expansion, which separates atoms with different momenta. However, to obtain $\omega_{B}$ using this method requires that the experiment be repeated for multiple hold times in order to map out the oscillations of the atomic momentum. The local acceleration due to gravity has been measured using this technique to a precision of around $10^{-7}$ in a experiment lasting about one hour \cite{Poli11}. The advantage of looking at the light leaking out of a cavity rather than measuring the momentum of the atoms directly is that a single continuous measurement can be performed. In contrast to a free-space optical lattice, inside a cavity, the light is strongly modified by the atoms, provided the conditions for large cooperativity $C$ are met, namely $C=\Omega_{0}^2/(2 \kappa \gamma) \gg 1$, where $2 \gamma$ is the atomic spontaneous emission rate in free space, $2 \kappa$ is the cavity energy damping rate and $\Omega_0 = d_{a} \sqrt{\frac{ \omega_c}{\hbar \epsilon_{0} V_{c}}}$ is the single photon Rabi frequency. In this latter expression, $\omega_{c}$ and $V_{c}$ are the cavity mode frequency and volume, respectively, and $d_{a}$ is the atomic dipole moment. We have previously predicted that this cavity-based technique should permit a measurement of the local acceleration due to gravity with a precision of around $10^{-6}$ in an experiment lasting only one \mbox{second \cite{Ven09}.} The disadvantage of working in a cavity is that quantum measurement backaction, in the form of random fluctuations in the cavity field due to photons spontaneously leaking out of the cavity, heats up the cold atoms and limits the coherence time of the measurement \cite{Corney98,Mekhov09,leroux10,Mekhov12,Niedenzu13,Lee14,Ell15,Lee15}. {The coherence time is particularly hard to calculate in the Bloch oscillating case \cite{Ven13} due to the time dependence introduced by the Bloch oscillations, especially in the presence of many particles, but it can be roughly estimated to be $\tau=\tau_{\mathrm{sp}}/(1+C)$ \cite{Ven09} at cavity resonance, where $\tau_{\mathrm{sp}}^{-1}=2 \gamma \vert \alpha \vert^2 \Omega_{0}^2/\Delta_{a}^2$ is the spontaneous emission rate at an antinode. The factors $\vert \alpha \vert^2$ and $\Delta_{a}$ are the mean number of cavity photons and the detuning of the laser from atomic resonance, respectively, and will be properly defined in the next section. The numerical value of $\tau$ for the parameters considered in this paper will be given in \mbox{Section \ref{sec:metrology}.} Of course, Bose--Einstein condensates can be continuously measured and used for sensing without a cavity, e.g., \cite{Ruostekoski98,Dalvit02,Saba05,Lee12,Java13}, but the cavity case is particularly interesting, because it allows for a strong atom-light interaction even in the quantum regime.}

The optomechanical cooling of a cavity mirror by dynamical backaction becomes most efficient in the resolved sideband regime $\omega_{m} > \kappa$, where $\omega_{m}$ is the angular frequency of the mechanical oscillator \cite{Schliesser08}. In this regime, the motion of the oscillator imprints sidebands on the cavity field at $\omega_{c} \pm \omega_{m}$, and if the cavity is pumped on the red sideband, the incident photons are resonantly up-converted to the cavity frequency at the expense of oscillator phonons. When the photons subsequently leak out of the cavity, they take this extra energy with them, thereby giving a cooling effect. If, on the other hand, the cavity is pumped on the blue sideband, then dynamical backaction leads instead to a parametric instability, where the amplitude of the mirror motion increases \cite{Braginsky+Manukin77}. Optical sideband cooling has also been successfully used to cool the collective motion of atoms in cavities \cite{Vul00,SchleierSmith11}. An intriguing question, therefore, is what happens to atomic Bloch oscillations in a cavity in the resolved sideband regime $\omega_{B} > \kappa$? In a recent paper, we addressed this problem and found that the backaction-induced modulation of the cavity field in a {ring} cavity at $\omega_{B}$ causes Bloch oscillating atoms to be transported up or down the optical lattice, depending on the \mbox{detuning \cite{Gol14}.} When the cavity is pumped on the red sideband, energy is extracted from the atoms, and they are coherently transported downhill (with the force), whereas when the cavity is pumped on the blue sideband, energy is deposited in the atomic motion, and they are coherently transported uphill. In our previous work, we explained this effect by analogy to the transport effects that can be generated in atom clouds under externally-imposed amplitude/phase modulation of a free-space optical \mbox{lattice \cite{Hal10,Kud11,Alb10,Tar12},} highlighting that in the Bloch-oscillating-atoms-in-a-cavity case, this was a self-induced effect. In the present paper, we develop an alternative explanation in terms of \mbox{cavity optomechanics.}

While the presence of two degenerate travelling wave modes in a ring cavity leads to rich possibilities for the Bloch oscillation dynamics, as we examined earlier in \cite{Gol14} (see \cite{Sam14} also), many of the key aspects, such as transport, are possible even in a standing wave cavity. In fact, as we noted in the supplement of \cite{Gol14}, many aspects of the Bloch oscillation dynamics in a ring cavity with equal driving power of the clockwise and anti-clockwise running wave modes can be captured by treating just one of the standing wave modes (the symmetric cosine) as dynamic and ignoring the other (antisymmetric sine mode) completely. Moreover, since we find that the analogy to an optomechanical system, which is a central goal of this work, is easier and clearer to present with a single standing wave cavity mode, we choose a Fabry--Perot cavity as the setting in this paper.

The plan for the paper is as follows. In Section \ref{sec:sec2}, we introduce the Hamiltonian and the mean field equations of motion describing the dynamics of a BEC confined inside a Fabry--Perot cavity. Following this, we present results from the numerical simulation of the mean-field equations of motion illustrating the typical dynamics in the regime $\omega_B \sim \kappa$. We find that the dynamics is particularly sensitive to the initial density distribution of the condensate with centre-of-mass oscillations and transport dynamics for an initial wave function spread out over many lattice sites and breathing dynamics when the initial state is localised. Moreover, we observe that the appearance of transport is correlated with an imbalanced strength in the sidebands at $\pm \omega_B$ of the cavity field's power spectral density. We also provide numerical results exemplifying the possibilities to control the transport velocity by changing the pump-cavity resonance detuning. Following the insights gained from the numerical analysis of the dynamics, in Section \ref{sec:sec3}, we develop a mapping between the Bloch-oscillating-BEC-in-a-cavity system to the standard ``cavity-mirror-on-a-spring'' optomechanical system; in Section \ref{sec:standard}, we compare and contrast the two systems, focusing on the difference between the Wannier--Stark ladder, which provides the single-particle spectrum for the ``Bloch oscillator'', and the harmonic oscillator spectrum, which occurs in the standard case. \mbox{In Section \ref{sec:transport}}, we use the optomechanical mapping to explain the coherent transport observed from the numerics in analogy with sideband cooling and amplification in cavity optomechanics \cite{Asp14}. \mbox{Apart from} providing a novel way to view the transport in the backaction modulated intracavity optical lattice, the mapping also provides quantitative analytical expressions that compare well to the numerical results presented in Section \ref{sec:sec2}. After providing some additional comments and discussion regarding the robustness of the Bloch frequency as the modulation frequency in the problem and its suitability for metrology in Section \ref{sec:metrology}, we conclude the paper in Section \ref{sec:conclusions}.

\section{Bloch Oscillation and Transport in a Cavity}\label{sec:sec2}

Our system of interest is a Bose--Einstein condensate trapped inside a standing wave optical cavity, as shown schematically in Figure \ref{fig:fig1}. We neglect dynamics along the transverse degrees of freedom (which is justified assuming strong external confinement), effectively reducing the dynamics to a single spatial dimension $z$. The cavity mode of resonance frequency $\omega_c$ is pumped through a lossless input-output coupling mirror by a laser with frequency $\omega_L=ck_r$, where $c$ is the speed of light in vacuum and $\hbar k_r$ is the recoil momentum. For the purposes of this paper, we consider the wavenumber of the cavity mode $k = \omega/c$ and the laser $k_r$ to be identical; this approximation will be examined in Section \ref{sec:metrology}. The light is detuned far enough from the atomic resonance that the excited state of the atoms can be adiabatically eliminated. For simplicity, we also ignore atomic collisions, which may be negligible in an experiment, either because the scattering cross-section is naturally small \cite{Alb09} or has been made small through the use of a tunable Feshbach resonance \cite{Hal10}. In a frame rotating at $\omega_L$ and in the dipole and rotating wave approximations, the Hamiltonian with the excited state adiabatically eliminated is given by:
\begin{align}
\hat{H} = -\hbar \left[ \Delta_c \adagger \aa + i(\eta^\ast\aa -\eta \adagger)\right]
+ \int \mathrm{d}z \psidagger \left(-\frac{\hbar^2}{2m}\,\partial_z^2 + \hbar U_0 \adagger \aa \cos^2(k_r z) - Fz \right)\!\ppsi
\label{eq:H}
\end{align}
where the annihilation operator $\aa$ corresponding to the cavity mode and $\ppsi(z)$ acting on the atomic field obey bosonic commutation relations. Here, $\Delta_c=\omega_L-\omega_c$ is the detuning of the driving laser from the bare cavity resonance frequency, and $\eta=\sqrt{J\kappa}$ represents the driving strength of the light mode for an incident flux of $J$ photons per unit time. In the far-detuned dispersive regime, the cavity light field provides a conservative lattice potential for the atoms with a spatial profile given by the standing wave mode $\cos^2(k_r z)$ and depth proportional to $U_0 = \Omega_0^2/\Delta_a$, which is a function of the pump's detuning from atomic resonance $\Delta_a = \omega_L-\omega_a$. $F<0$ is the uniform and constant bias force that drives the Bloch oscillations. In Equation (\ref{eq:H}), the term representing the intracavity optical lattice is also the dispersive interaction term between the atoms and the light field, \emph{i.e.},
\begin{eqnarray}
\label{eq:Hi}
\hat{H}_{\mathrm{disp }}&=& \hbar U_0 \n\,\mathcal{C}
\end{eqnarray}
where $\n= \adagger \aa $ is the total number of photons and $\mathcal{C}[\ppsi]=\langle\cos^2(k_rz)\rangle$ is the overlap between the atomic density and the optical mode that characterizes the degree of spatial ordering of the atoms. When combined with the free evolution term for the cavity $-\hbar \Delta_c \n$, the interaction term may be viewed as a dynamical shift to the effective cavity resonance frequency proportional to $\mathcal{C}[\ppsi]$. In addition, Equation (\ref{eq:Hi}) gives us a glimpse of what to expect from the dynamics: the Bloch oscillation dynamics of the atomic state causes $\mathcal{C}[\ppsi]$ to oscillate, whose coupling to the intracavity photon number in turn leads to a modulation of the intracavity lattice depth \cite{Ven09,Ped09}.

To solve the full nonlinear dynamics in the mean-field limit, we write the Heisenberg--Langevin equations, $i\hbar\partial_t\aa =[\aa,\hat{H}]-i\hbar\kappa\,\aa$ and $i\hbar\partial_t\ppsi=[\ppsi,\hat{H}]$, ignoring all input noise operators, whose means are zero. We also neglect atom losses over the time scales of interest, so that $N = \langle\int dz \hpsid(z)\hpsi(z)\rangle$ is constant. Letting $\alpha =\langle\aa \rangle$ and $\psi=\langle\ppsi\rangle/\sqrt{N}$ and factoring the expectation values of operator products, we obtain the equations of motion in the mean-field approximation,
\begin{eqnarray}
\label{eq:alpha1}
\partial_t\,\alpha &=& -(\kappa-i\Delta_f)\,\alpha + \eta \\
\label{eq:psi}
i\hbar\,\partial_t\,\psi &=& \left(-\frac{\hbar^2}{2m}\,\partial_z^2+\hbar U_0 \had \ha \cos^2(k_r z) - Fz\right)\psi \
\end{eqnarray}

The quantity $\Delta_f= \Delta_c- N U_0 \mathcal{C}$ is the total effective detuning and includes both the drive-cavity detuning and the Stark shift due to the atom-light coupling. Again, we see that the lattice modulation is driven by changes in $\mathcal{C}$ during Bloch oscillations via the effective detuning $\Delta_f$.

We now consider the results of a numerical solution of the coupled Equations (\ref{eq:alpha1}) and (\ref{eq:psi}). Throughout, we choose $^{88}$Sr atoms accelerating under gravity in a 689~nm intracavity lattice \mbox{(\emph{i.e.}, lattice} spacing $d=\pi/k_r=344.5$~nm). In this case, we have $\omega_B = 2\pi\times 745$~Hz, and the recoil frequency $\omega_r\equiv\hbar k_r^2/(2m)=2\pi\times 4.78$~kHz. As discussed in detail in \cite{Gol14}, transport dynamics is observed only in the limit $\omega_B \sim \kappa$, as this allows for a finite phase lag between the atomic dynamics and the cavity field response that gives rise to the modulation of the intracavity lattice \cite{Hal10,Kud11}. We therefore stick to this regime throughout this paper and set $\kappa = 2 \pi \times 1$ kHz, which is close to the value achieved in the experiment by Wolke \emph{et al.}\ \cite{Wolke2012}. {Figures \ref{fig:fig2}} and \ref{fig:fig3} show the typically-observed atomic and lattice amplitude dynamics for two extremal possibilities for the initial atomic distribution. \scalebox{.95}[1.0]{In Figure \ref{fig:fig2a},} coherent backaction-induced transport dynamics is evident from the behaviour of the centroid of an initial atomic wave packet spread over more than 20 lattice sites. In addition to periodic dynamics that repeats every Bloch period, the centroid also drifts either uphill ($\Delta_c-NU_0 \mathcal{C}>0$) or downhill ($\Delta_c-NU_0 \mathcal{C}<0$). On the other hand, when the initial atomic wave packet
is localised within a single lattice site, there is no translation of the centroid (see Figure \ref{fig:fig2b}), and the atomic wave packet undergoes periodic breathing dynamics. Transport dynamics may be understood as a direct consequence of the atomic backaction-induced modulation of the intracavity lattice depicted by the red and blue curves in Figure \ref{fig:fig3} \cite{Gol14,Tho02,Alb09,Hal10,Kud11}. Both the modulation and the phase lag between the lattice dynamics and atomic Bloch oscillation can be controlled by tuning $\Delta_c$, allowing for uphill, as well as downhill transport. In contrast, in the case of a localised initial atomic wave function, the lattice amplitude modulation is suppressed (black curve in Figure \ref{fig:fig3}). As a result, some of the interesting features, such as coherent delocalisation \cite{Tar12}, expected for localised initial wave packets in amplitude modulated lattices are absent in our case, and the atomic dynamics observed is similar to what is expected in a stationary tilted optical lattice.
\vspace{-12pt}
\begin{figure}[H]
\centering
 \subfloat[]{\label{fig:fig2a}
\includegraphics[width=0.42\textwidth,height=8.3cm]{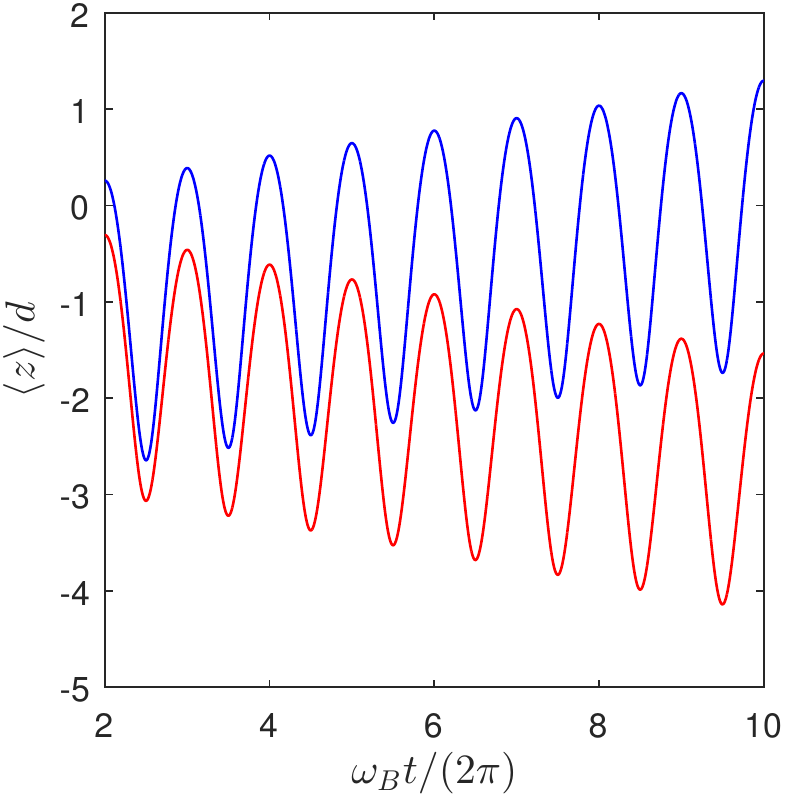}}
 \hfill
 \subfloat[]{\label{fig:fig2b}\includegraphics[width=0.54\textwidth,height=8.6cm]{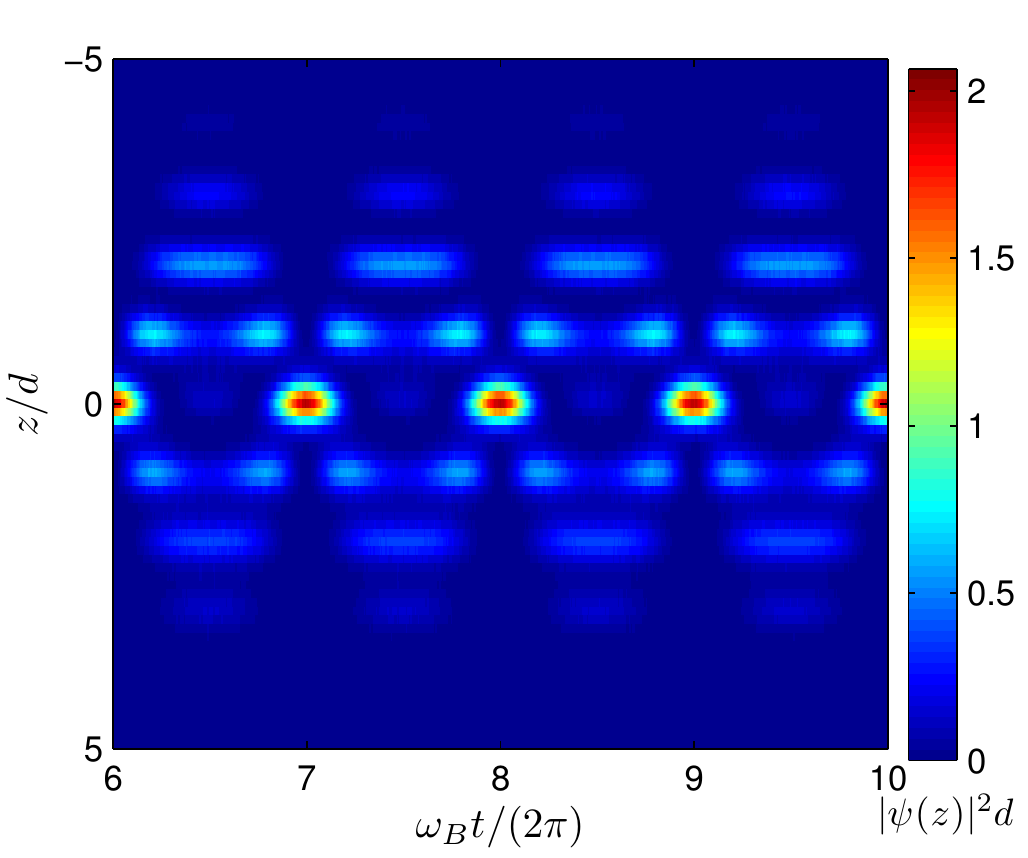}} 
\caption{Backaction-induced atomic transport and breathing dynamics in a cavity with \mbox{$U_0 = - 2 \pi \times 1$ Hz,} $\kappa = -NU_0 = 2 \pi \times 1$ kHz and Bloch frequency $\omega_B = 2 \pi \times 744.5$ Hz. \mbox{(\textbf{a}) Condensate} centroid position as a function of time showing uphill (blue, $\Delta_c-NU_0=1.3\kappa$) and downhill (red, $\Delta_c-NU_0=-0.7\kappa$) transport for an initial atomic wave packet delocalised over \mbox{20 lattice} sites; (\textbf{b}) breathing dynamics of the condensate density at the Bloch period for \mbox{$\Delta_c-NU_0=-0.7\kappa$} and an initial atomic wave packet localised within one lattice site. }
\label{fig:fig2}
\end{figure}

\begin{figure}[H]
\centering
\includegraphics[width = 0.49\textwidth,height=8.6cm]{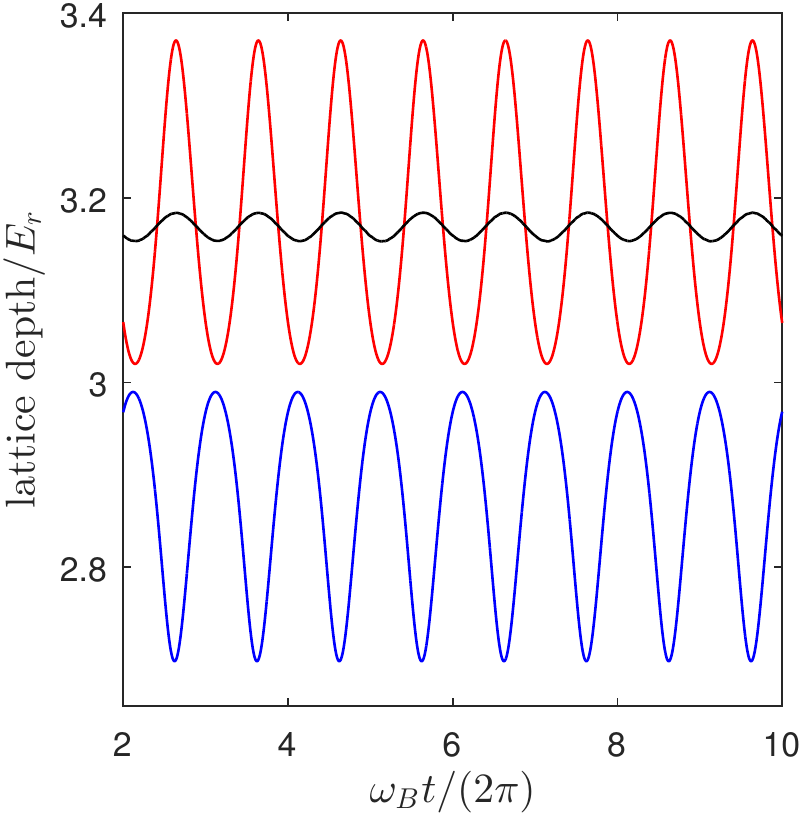}
\caption{Backaction-induced modulation of the lattice depth (in units of the recoil energy \mbox{$E_r=\hbar\omega_r$)} as a function of time during intracavity Bloch oscillations. The blue (red) curve corresponds to parameters giving uphill (downhill) transport in Figure \ref{fig:fig2a}. The black curve corresponds to the breathing dynamics plotted in Figure \ref{fig:fig2b}.}
\label{fig:fig3}
\end{figure}

In this paper, we develop another way to understand the transport behaviour observed in \mbox{Figure \ref{fig:fig2a}. }This is motivated in part by the behaviour of the power spectral density of the intracavity light field $\alpha(t)$ shown in Figure \ref{fig:fig4a}; we see that, as expected from the field modulation, there are sidebands at $\pm \omega_B$ and its higher harmonics. More interestingly, there is a marked asymmetry in the strength of the sidebands, reminiscent of cavity sideband cooling \cite{Asp14}, with uphill (downhill) transport corresponding to larger strength at $- \omega_B$ ($+\omega_B$) and its harmonics. While in the following section, we develop a formal mapping to an optomechanical system to clearly explain this behaviour of the spectrum, the motivation for such a mapping may be gleaned from Figure \ref{fig:fig4b}. There, we plot the expectation value of the force applied by the (dynamic) intracavity lattice on the atomic wave packet $\langle F_{\mathrm{lattice}}\rangle = - \int dz \, \vert \psi(z,t) \vert^2 \sin(2k_rz) U_0 \vert \alpha(t) \vert^2$ as a function of the mean atomic displacement, \emph{i.e.}, the centroid position $\langle z \rangle$. For $\Delta_f<0$ ($>0$), during each Bloch period, the system traverses a anti-clockwise (clockwise) closed loop in the $\langle F_{\mathrm{lattice}} \rangle$--$\langle z \rangle$ plane, giving rise to a net negative (positive) work done on the atoms proportional to the area of the loop. This excess work has to be compensated for by the light. Hence, as depicted in the inset of Figure \ref{fig:fig4b}, we can associate the positive (negative) work with the case where the driving laser frequency $\omega_L$ is blue (red) detuned with respect to the cavity resonance $\omega_c$, and to enter the cavity, a laser photon has to give up (absorb) energy in units of the Bloch frequency $\omega_B$ enabled by the interaction with atoms. Naturally, such preferential up or down conversion leads to an asymmetric strength of the sidebands at harmonics of $\omega_B$ in the cavity field spectrum.

Another interesting result that we find empirically by performing numerical simulations of the coupled dynamics over a wide range of parameters is that the dominant oscillation frequency of observables is very robustly fixed to the Bloch frequency. We do not observe any systematic shift from the Bloch frequency due to the dynamical backaction. This is especially significant in the context of the mapping to an optomechanical system that is pursued in the next section, since in conventional optomechanical systems, the radiation pressure coupling-induced dynamical backaction causes a shift in the oscillation frequency of the mechanical element, the so-called optical spring\mbox{ effect \cite{Asp14}.} \mbox{In Section \ref{sec:standard}}, we will show analytically that such an optical spring effect is absent for our situation.

\begin{figure}[H]
\centering
 \subfloat[]{\label{fig:fig4a}
\includegraphics[width=0.45\textwidth,height=8.6cm]{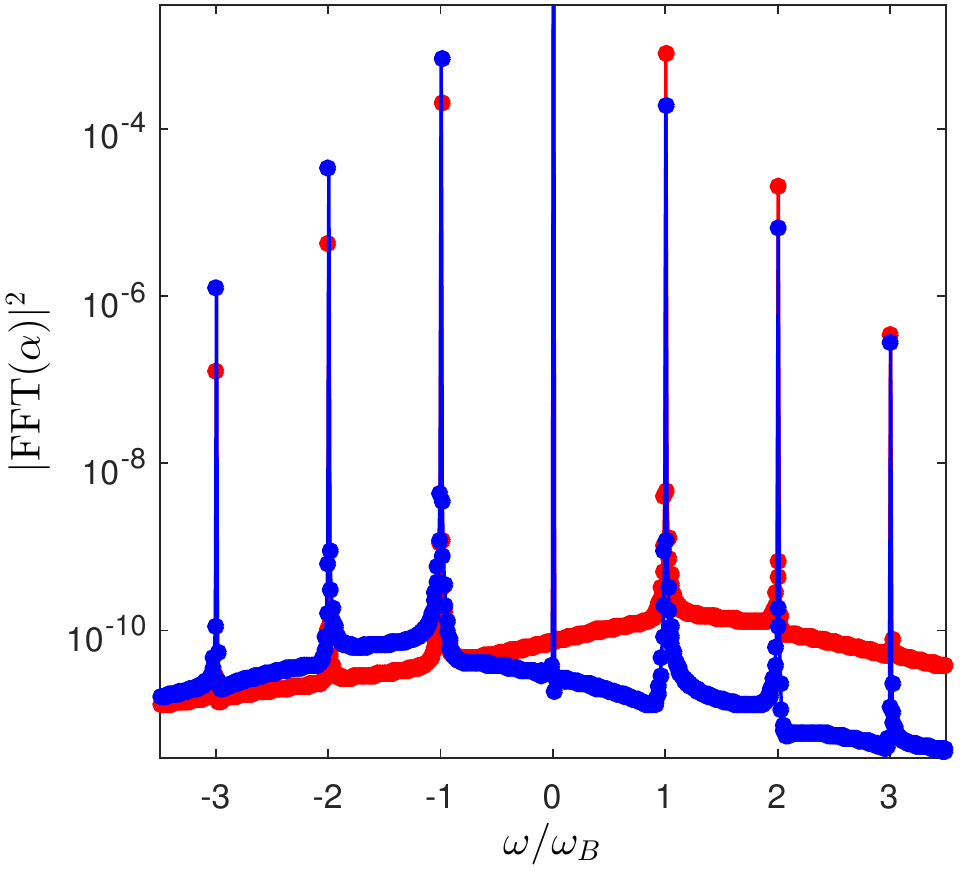}}
 \hfill
 \subfloat[]{\label{fig:fig4b}\includegraphics[width=0.49\textwidth,height=8.6cm]{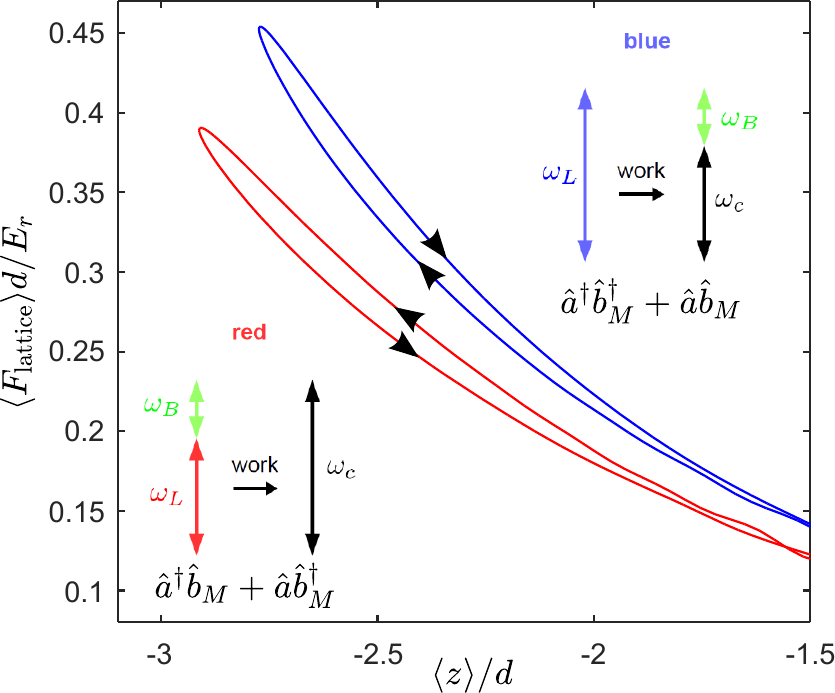}} 
\caption{Power spectral density of the cavity field and force-displacement plots for the atomic dynamics. The parameters are the same as in Figure \ref{fig:fig2a}. (\textbf{a}) Power spectral density of $\alpha(t)$, obtained by a fast Fourier transform at resolution $\omega_{B}/100$, showing asymmetric sidebands at multiples of $\pm \omega_B$ with more power in the red upper (blue lower) side band corresponding to downhill (uphill) transport in Figure \ref{fig:fig2a}. Note that the frequency origin corresponds to the driving laser frequency, which is different for the red and blue points. (\textbf{b}) Average of the force applied by the dynamical intracavity optical lattice as a function of the atomic wave packet centroid position. During uphill (downhill) transport of the atoms, the blue (red) curve is traversed in the clockwise (anti-clockwise) direction, corresponding to positive (negative) work done on the atoms and accounted for by the down-conversion (up-conversion) of the blue-detuned (red-detuned) pump laser photons (see the inset). The insets also show the dominant terms in the effective optomechanical Hamiltonian, which we derive in Section 3, with the operator $\hat{b}_M$ annihilating a quantum of excitation from the Bloch oscillator. The area enclosed by the loops gives the work done on the atoms by the lattice. }
\label{fig:fig4}
\end{figure}

\section{Mapping to an Optomechanical Hamiltonian} \label{sec:sec3}

The eigenstates of a quantum particle in a periodic lattice of finite extent that is tilted by an additional linear potential are the Wannier--Stark (WS) states \cite{Wan60,Tho02}. We define them as the solutions $\varphi_n(z)$ of:
\begin{align}
\left [-\partial_z^2 + s_0 \cos^2 \left(z\right) - f z \right ] \varphi_n (z) = (e_0 + n \omega_B) \varphi_n(z) \label{eq:WSdefn}
\end{align}

In order to make all quantities dimensionless, we have multiplied the position coordinate $z$ by the recoil momentum $k_r$ and scaled energies by the recoil energy $E_r$, but retained their original symbols. Thus, $f = -\omega_B/\pi$, with $\omega_B>0$ being the Bloch frequency scaled by the recoil frequency. $s_0$ denotes a fixed (time independent) lattice depth, and its relation to the cavity parameters will be explained below. The state $\varphi_n(z)$ is localised around the $n$-th lattice site; states at different lattice sites are simply related by a discrete translation of the coordinate, \emph{i.e.}, $\varphi_{n+m}(z) = \varphi_n(z-m\pi)$, and we shall only include the states of the first band, ignoring higher bands (which are typically not excited during Bloch oscillations, which are an adiabatic phenomenon). The eigenspectrum $e_0 + n \omega_B$ is organized into a discrete ladder of states with an equal spacing between the rungs given by the Bloch frequency. Without loss of generality, we can set the reference energy $e_0 = 0$. The WS states will play an important role in the rest of this section, where we construct the mapping to an optomechanical system.

The coupled equations of motion (\ref{eq:alpha1}) and (\ref{eq:psi}) for the light and matter can be expressed in the dimensionless notation as:
\begin{align}
\dot{\alpha}(t) &= - (\kappa - i \Delta_f) \alpha(t) + \eta \label{eq:alphac}\\
i \partial_t \psi(z,t) &= \left(-\partial_z^2 + U_0 \vert \alpha (t) \vert^2 \cos^2(z) - fz \right) \psi(z,t) \label{eq:psiscaled}
\end{align}
where frequencies, such as $U_0$, have been scaled by $\omega_r$ and time by $\omega_r^{-1}$. As above, the same symbols are retained for the dimensionless variables and parameters as their dimensionful versions. We write the cavity field amplitude $\alpha(t)$ in terms of a time-independent and a time-modulated part:
\begin{align}
\alpha(t) = \alpha_{0} + \Delta \alpha(t) \label{eq:alphacres}
\end{align}
and use $\alpha_{0}$ to fix the WS basis $\braket{z}{\varphi_n} = \varphi_n(z)$, with the lattice depth $s_0 = U_0 \vert \alpha_0 \vert^2$ in Equation (\ref{eq:WSdefn}). This splitting also means that the intracavity photon number can be written as \mbox{$n(t) = \vert \alpha(t)\vert^2 = \vert \alpha_0 \vert^2 + \Delta n(t)$} with:
\begin{align}
\Delta n (t) = \alpha_0^{*} \Delta \alpha(t) + \alpha_0 \Delta \alpha^{*}(t) + \vert \Delta \alpha \vert^2(t) \label{eq:photfluct}
\end{align}

Next, we wish to expand the atomic wave function in the WS basis as (in Dirac notation):
\begin{align}
\ket{\psi(t)} = e^{-i \vartheta(t)} \sum_n c_n(t) \ket{\varphi_n} \label{eq:psiexpand}
\end{align}
where $\vartheta(t)$ is a site-independent dynamical phase factor that will be chosen briefly 
 in order to simplify the equations, and the wave function is normalized so that $\langle \psi(t) \vert \psi(t) \rangle =1$. The effective detuning $\Delta_f = \Delta_c - NU_0 \mathcal{C}$ plays a central role in the equations of motion (\ref{eq:alphac}) and (\ref{eq:psiscaled}), and hence, it is useful to evaluate the expectation value $\mathcal{C}$ in terms of the WS basis expansion as:
\begin{align}
\mathcal{C} \equiv \bra{\psi(t)}\cos^2(z)\ket{\psi(t)} &= \sum_{nm} c_m^{*} c_n \bra{\varphi_m} \cos^2(z) \ket{\varphi_n} \approx \gamma_0 + \gamma_1 \left ( \sum_n c_n^{*}c_{n+1}(t) + h.c. \right) \label{eq:cos2xWS}
\end{align}
with $\gamma_0 = \bra{\varphi_n} \cos^2(z) \ket{\varphi_n}$ and $\gamma_1 = \bra{\varphi_n} \cos^2(z) \ket{\varphi_{n+1}}$, where we have used the following property for the translation of the WS states:
\begin{align*}
\braket{x}{\varphi_{n+m}} = \braket{x-n\pi}{\varphi_{m}}
\end{align*}

Note that in the above approximate expression, we neglect the overlap {beyond} nearest neighbour WS states, which is justified when working at mean lattice depths $s_0$ large enough that the WS states are well localised. The matrix elements $\gamma_{0}$ and $\gamma_{1}$ are key parameters that appear in the equations of motion. In particular, the static part of the field obeys:
\begin{align}
\alpha_0 = \frac{\eta_c}{\kappa - i \delta_0} \label{eq:alphac0eqn}
\end{align}
where $\delta_0 = \Delta_c - NU_0 \gamma_0$, and upon substituting Equations (\ref{eq:alphacres}) and (\ref{eq:psiexpand}) into the equations of \mbox{motion (\ref{eq:alphac})} and (\ref{eq:psiscaled}), we obtain:
\begin{align}
\frac{d \Delta \alpha(t)}{dt} &= ( -\kappa + i \delta_0) \Delta \alpha (t) - i U_0 \gamma_1 (\alpha_0+\Delta \alpha(t)) \sum_n \left(d_n d_{n+1}^{*}(t) + d_n^{*}d_{n+1}(t) \right) \label{eq:delalphaeq}\\
i \dot{d}_n(t)&= n \omega_B d_n(t) + U_0 \gamma_1 \Delta n(t) \left [ d_{n+1}(t) + d_{n-1}(t) \right ] \label{eq:WSbasiseq}
\end{align}
where we have introduced the scaled WS amplitudes $d_n(t) = \sqrt{N} c_n(t)$, so that the average atom occupation number at site $n$ is given by $\vert d_n(t) \vert^2$. In arriving at this form of the equations of motion, we used our freedom to choose the global phase $\vartheta(t)$ by setting it equal to $\vartheta(t) = \int_0^t dt^{'} \, U_0 \, \gamma_0 \, \Delta n(t^{'})$.

It is straightforward to show that Equations (\ref{eq:delalphaeq}) and (\ref{eq:WSbasiseq}) can be derived from the following effective (and fully quantized) Hamiltonian:
\begin{align}
\hat{H}_{\mathrm{eff}} = - \delta_0 \hbd \hb + \sum_{n} n \omega_B \hdd_n \hd_n + U_0 \gamma_1 \hat{\Delta n} \sum_n \left(\hdd_n \hd_{n+1} + \hdd_{n+1} \hd_n \right) \label{eq:Heffv1}
\end{align}
where $\langle \hb \rangle(t) = \Delta \alpha (t)$ and $\hat{\Delta n}(t) = \alpha_0^* \hb + \alpha_0 \hbd + \hbd \hb$. The collective atomic operator $\hd_n = \ket{0}\bra{\Psi_{N,n}}$, where $\ket{\Psi_{N,n}}$ represents the state with $N$ atoms coherently occupying the WS state at site $n$. This immediately ensures the following commutators for the atomic operators:
\begin{align}
\left[ \hd_n,\hdd_m \hd_m \right] &= \delta_{nm} \hd_n \nonumber \\
\left[ \hdd_n, \hdd_m \hd_m \right ] &= -\delta_{nm} \hdd_n \label{eq:atomcomms}
\end{align}

These commutators, along with the standard bosonic commutators for $\hb$, can be used to check that the mean field Equations (\ref{eq:delalphaeq}) and (\ref{eq:WSbasiseq}) follow from the Heisenberg equations of motion corresponding to the Hamiltonian Equation (\ref{eq:Heffv1}). In this representation, the average $\langle \hdd_n \hd_n \rangle \equiv \vert d_n \vert^2$ gives the occupation number of the $n$-th WS ladder state. The multi-site operators:
\begin{align}
\hb_M &= \sum_n \hdd_n \hd_{n+1} = \sum_n \ket{\Psi_{N,n}}\bra{\Psi_{N,n+1}} \label{eq:aopbloch}\\
\hn_M &= \sum_n n \hdd_n \hd_n = \sum_n n \ket{\Psi_{N,n}} \bra{\Psi_{N,n}} \label{eq:nopbloch}
\end{align}
define a ``Bloch oscillator'', analogous to the mechanical oscillator in usual optomechanical systems, and allows us to rewrite Equation (\ref{eq:Heffv1}) in the form of an optomechanical Hamiltonian:
\begin{align}
\hat{H}_{\mathrm{optomech} \, \mathrm{Bloch}} & = - \delta_0 \hbd \hb + \omega_B \hn_M + U_0 \gamma_1 \hat{\Delta n} (\hb_M + \hbd_M) \\ & \approx - \delta_0 \hbd \hb + \omega_B \hn_M + U_0 \gamma_1 \vert \alpha_{0} \vert (\hb+\hbd) (\hb_M + \hbd_M) \ \label{eq:blochoscillator}
\end{align}
where in the second line, we made the approximation $\hat{\Delta n}(t) \approx \vert \alpha_0 \vert (\hb + \hbd)$.

\section{Comparison to Standard Cavity Optomechanics}
\label{sec:standard}

The standard optomechanical Hamiltonian is generally taken to be:
\begin{equation}
\hat{H}_{\mathrm{optomech} \, \mathrm{standard}} =- \Delta \, \hbd \hb + \Omega_{M} \hat{c}^{\dag} \hat{c}- g_{0} \vert \alpha_{0} \vert (\hb +\hbd)(\hat{c}+\hat{c}^{\dag}) \ \label{eq:standardH}
\end{equation}

This is Equation (28) in the review on cavity optomechanics by Aspelmayer \emph{et al}. \cite{Asp14} adapted to our notation. The operators $\hb$ and $\hbd$ are for the cavity field as before, but $\hat{c}$ and $\hat{c}^{\dag}$ are for a harmonic oscillator of frequency $\Omega_{M}$, e.g., a mirror on a spring. The third term in the Hamiltonian accounts for the radiation pressure on the mirror, where $g_{0}$ is the coupling rate per photon. Specifically, the force on the mirror can be written as $\hat{F}=(\hbar g_{0}/ Z_{\mathrm{ZPF}} ) \hbd \hb$, where $Z_{\mathrm{ZPF}}=\sqrt{\hbar/(2 m_{\mathrm{eff}} \Omega_{M})}$ is the amplitude of zero-point fluctuations of the mirror position, $m_{\mathrm{eff}}$ being the effective motional mass of the mirror-spring system. The detuning $\Delta=\Delta_{c}-2 g_{0}^2 \vert \alpha_{0} \vert^2 / \Omega_{M}$ is the bare laser-cavity detuning minus the frequency shift due to the displacement in the mirror's average position caused by the static part of the radiation pressure.

As can been seen, the Bloch oscillator Hamiltonian Equation (\ref{eq:blochoscillator}) replicates the standard optomechanical Hamiltonian Equation (\ref{eq:standardH}) term by term. The role of the coupling constant $g_{0}$ is played by $U_0 \gamma_1$, and comparing the oscillator energies (second term in each Hamiltonian), we find that we can associate the total excitation energy of the mechanical harmonic oscillator with the total gravitational (or equivalent) potential energy of the atoms distributed across the ladder of WS states. However, there are also some basic differences between the two systems (see Figure \ref{fig:fig1}). The first thing to notice is that the expression for the annihilation operator for a harmonic oscillator in terms of its eigenstates $\ket{n}$ is:
\begin{align*}
\ha = \sum_n \sqrt{n} \ket{n} \bra{n+1} \
\end{align*}

This is different from the equivalent expression for the Bloch oscillator in terms of the WS states, as given in Equation (\ref{eq:aopbloch}), where there are no $\sqrt{n}$ factors in the sum describing $\hat{b}_M$. In fact, the origin of the $\sqrt{n}$ factors for a harmonic oscillator is the assumption of a normalized reference or vacuum state, which does not exist for Bloch oscillations, at least in an infinite lattice. This highlights the point that the annihilation operator for the Bloch oscillator is really a shift operator, and there is no preferred reference state along the tilted lattice. Further differences and similarities to the usual harmonic oscillator can be identified by writing down the relevant commutators for the Bloch oscillator:
\begin{align}
\left[ \hb_M,\hn_M \right] = \hb_M, & \,\, \left[ \hbd_M,\hn_M \right] = -\hbd_M \label{eq:bocomm1}\\
\left[ \hb_M, \hbd_M \right] = 0 \label{eq:bocomm2}
\end{align}
and moreover, $\hbd_M \hb_M \neq \hn_M$.

The above commutators can be used to write down the equation of motion for the quantity $\langle \hb_M + \hbd_M \rangle$, which serves as the analogue to the mirror position, as:
\begin{align}
\frac{d^2 \langle \hb_M + \hbd_M \rangle}{dt^2} = -\omega_B^2 \langle \hb_M + \hbd_M \rangle \label{eq:nospring}
\end{align}

Crucially, this tells us that the dynamics of the ``position'' operator are unaffected by the interaction, unlike the radiation pressure coupling in typical optomechanical setups \cite{Asp14}. Even in the presence of the backaction, the oscillation frequency of the system is unchanged from $\omega_B$, \emph{i.e.}, {there is no optical spring effect}, and the Bloch frequency is robust as observed in Section \ref{sec:sec2}. In hindsight, this robustness is a consequence of the assumptions of our model, where the dynamical backaction from the Bloch oscillating atoms only modulates the {amplitude} of the intracavity lattice, leaving the light wavelength unchanged, despite its coupling to the atoms. This means that the Bloch frequency is completely unaffected by the backaction, because it depends only on the wavelength of the light and is independent of the lattice depth. We examine the validity of this assumption and its consequences for metrology further in Section \ref{sec:metrology}.

\section{Transport as a Manifestation of Cavity Amplification or Cooling}
\label{sec:transport}

The ability to cool/amplify the motion of a mechanical oscillator by coupling it to an optical resonator via radiation pressure is one of the most important developments in optomechanics. \mbox{The underlying} mechanism is most easily understood in the so-called resolved sideband regime, where $\omega_M \geq \kappa$ (with $\omega_M$ denoting the mechanical element's frequency). As depicted in the inset of Figure \ref{fig:fig4b}, cooling (amplification) results when the cavity is pumped at a frequency red (blue) detuned from its resonance requiring an up (down) conversion of the photon's energy in order to enter the cavity with the energy difference extracted from (supplied to) the mechanical element's motional energy. This energy transfer is also apparent in the power spectrum of the cavity light field shown in Figure \ref{fig:fig4a} as an asymmetry in the amplitude of the sidebands due to the preferential scattering occurring during the frequency up or down conversion.

The amplification and cooling of the mechanical oscillator by the dispersive coupling to a light mode can also be understood using a classical argument. The finite ring-down time of the cavity $\kappa^{-1}\sim \omega_M^{-1}$ implies a time lag between the change in the position of the cavity mirror (see schematic Figure \ref{fig:fig1}b) and the intracavity photon number $\had \ha$. Since the force on the mirror is proportional to the photon number, the force falls below or above its stationary value depending on whether the direction of motion of the mirror brings the cavity closer to or further away from the cavity resonance ({{\mbox{the stationary} value of the force is calculated by assuming that the cavity field instantaneously responds to the changes in its length due to the moving mirror \emph{i.e.}, in the $\kappa \gg \omega_M$ limit}}). \mbox{When the} mirror dynamics are plotted in a force \textit{versus} displacement diagram, one therefore expects loops whose finite area $\oint \vec{F}(x) \cdot \vec{dx} $ gives positive (negative) work done on the mirror when the cavity is blue (red) detuned from the driving laser. 

In the numerical solutions presented in Figure \ref{fig:fig2a}, we see clear evidence for coherent directed transport up or down the tilted lattice in the regime with $\omega_B \sim \kappa$. This motion is non-conservative, because the atomic potential energy increases/decreases steadily as a function of time. Moreover, Figure \ref{fig:fig4b} shows that the average force on the atoms as a function of the centroid of the atomic cloud makes loops of finite area, as expected. Thus, it is interesting to ask if the transport process can be understood in a manner analogous to sideband cooling/amplification. The centroid position is given in terms of the WS basis coefficients as \cite{Tho02}:
\begin{align}
\langle z \rangle = Z_0 + d \sum_n n \vert d_n (t) \vert^2 + Z_1 \sum_n \left(d_n^{*} d_{n+1} e^{-i \omega_B t} + c.c. \right) \label{eq:zposition}
\end{align}
with $Z_0 = \bra{\varphi_n} z \ket{\varphi_n}$ and $Z_1 = \bra{\varphi_n} z \ket{\varphi_{n+1}}$. The third term (proportional to $Z_1$) in Equation (\ref{eq:zposition}) describes the repeated oscillatory motion seen in Figure \ref{fig:fig2a}. The transport dynamics arises then from the second term, which is proportional to $\langle\hat{n}_M\rangle/N$. To that end, we examine the dynamics of the variable $\langle\hat{n}_M\rangle/N$:
\begin{align}
\frac{1}{N}\frac{d \qavg{\hn_M}}{dt} = iU_0\gamma_1 \Delta n(t) \frac{\qav{\hb_M-\hbd_M}}{N}
\end{align}

Noting that, $d\qav{\hb_M + \hbd_M}/dt = -i \omega_B \qav{\hb_M - \hbd_M}$, we can use the exact solution to Equation (\ref{eq:nospring}) given by:
\begin{align}
\langle \hb_M + \hbd_M \rangle (t) = 2 N \vert \sigma_1 \vert \cos(\omega_B t + \theta_1) \label{eq:cohsol}
\end{align}
with the initial mean-field state's site-to-site coherence $\langle \hb_M \rangle (t=0) = \sum_{n} (d_n^* d_{n+1})(t=0) = \sigma_1 e^{i \theta_1}$ with $\sigma_1$ real to write:
\begin{align}
\frac{1}{N}\frac{d \qavg{\hn_M}}{dt} = 2 U_0\gamma_1 \sigma_1 \sin(\omega_B t + \theta_1) \Delta n(t) \label{eq:centroiddyn}
\end{align}

Thus, the time rate of change of the mean position is directly proportional to the time-dependent part of the cavity photon number $\Delta n (t)$. In order to determine this function, we can first exactly solve Equation (\ref{eq:delalphaeq}), which now reads as:
\begin{align*}
\frac{d \Delta \alpha(t)}{dt} &= \left[ -\kappa + i (\delta_0 - 2NU_0\gamma_1 \sigma_1 \cos(\omega_B t + \theta_1))\right] \Delta \alpha (t) - i 2NU_0\gamma_1 \sigma_1 \cos(\omega_B t + \theta_1) \alpha_0
\end{align*}
in light of Equation (\ref{eq:cohsol}). For $t \gg \kappa^{-1}$, the above equation can be solved using the Jacobi--Anger expansion as, 
\begin{align}
\Delta \alpha(t) = -\alpha_0 e^{-i\frac{u_1}{\omega_B} \sin(\omega_B t + \theta_1)} \sum_n \frac{in\omega_BJ_{n} \left( \frac{u_1}{\omega_B}\right)e^{in(\omega t + \theta_1)}}{\kappa - i \delta_n}\label{eq:delalphasol}
\end{align}
where $u_1 = 2NU_0\gamma_1 \sigma_1$, $\delta_n = \delta_0 - n \omega_B$, and $J_n$ denotes a Bessel function of order $n$. {{Technically, the Jacobi--Anger} expansion \scalebox{.95}[1.0]{$e^{iy\sin(\beta)} = \displaystyle \sum_{n=-\infty}^{\infty} J_n(y)e^{in\beta}$} requires the number of lattice sites to be infinite, but we expect that the situation will not change qualitatively in a large, but finite box}.

 Our aim is to use the above solution Equation (\ref{eq:delalphasol}) to evaluate $\Delta n(t) = \alpha_0^* \Delta \alpha(t) + \alpha_0 \Delta \alpha^*(t) + \vert \Delta \alpha(t) \vert^2$. Before that, a few simplifying approximations are in order. For the parameter regimes of interest in this paper, $u_1/\omega_B < 1$, and in this limit, the Bessel functions $J_n$ quickly decrease in magnitude as $n$ increases. Thus, we can restrict the sum in Equation (\ref{eq:delalphasol}) to just $n = \pm 1$. With this approximation, we find:
\begin{align}
\vert \Delta \alpha(t) \vert^2 \approx \vert \overline{\Delta \alpha(t)} \vert^2 = \vert \alpha_0 \vert^2 J_1\left(\frac{u_1}{\omega_B}\right)^2 \omega_B^2 \left[\frac{1}{\kappa^2+\delta_1^2} + \frac{1}{\kappa^2+\delta_{-1}^2}\right] \label{eq:dclight}
\end{align}
so that at lowest order, the contribution to photon number is time independent. On the other hand, there is also an oscillatory component coming from:
\begin{align}
\alpha_0^* \Delta \alpha(t) + \alpha_0 \Delta \alpha^*(t) = i \omega_B \vert \alpha_0\vert^2 J_0\parlr{\frac{u_1}{\omega_B}} J_1\parlr{\frac{u_1}{\omega_B}}& \left [ e^{i(\omega_B t+\theta_1)}\parlr{\frac{1}{\kappa+i\delta_{-1}}-\frac{1}{\kappa-i\delta_1}} \right. \label{eq:aclight}\\ &\left . + e^{-i (\omega_B t+ \theta_1)} \parlr{\frac{1}{\kappa+i\delta_{1}}-\frac{1}{\kappa-i\delta_{-1}}} \right ] \nonumber
\end{align}
where in the substitution of Equation (\ref{eq:delalphasol}), we have approximated the exponential $e^{-iu_1 \sin(\omega_Bt+\theta_1)/\omega_B} \approx J_0(u_1/\omega_B)$. It is now key to observe from Equation (\ref{eq:delalphasol}) that $1/(\kappa - i \delta_{\pm 1})$ are the amplitudes of the sidebands at $\pm \omega_B$ of the intracavity light field. Substituting the expression Equations (\ref{eq:dclight}) and (\ref{eq:aclight}) in (\ref{eq:centroiddyn}), we obtain:
\begin{align}
\frac{1}{N}\frac{d \qav{\hn_M}(t)}{dt} = c_1 + 2U_0\gamma_1 \sigma_1 \vert \overline{\Delta \alpha(t)} \vert^2 \sin(\omega_B t + \theta_1) + h.h. \label{eq:centroidfinal}
\end{align}
where $h.h.$ denotes terms that are higher harmonics of $\omega_B$. As expected, there is a constant drift term $c_{1}$, which can be related to the transport velocity, which we define as the net drift of the centroid per Bloch period in units of the lattice spacing $d$:
\begin{align}
v_t/\pi = 2 \pi c_1/\omega_B &= U_0 \vert \alpha_0 \vert^2 \gamma_1 \sigma_1 J_0 \parlr{\frac{u_1}{\omega_B}} J_1 \parlr{\frac{u_1}{\omega_B}}2\kappa \sqlr{\frac{1}{\kappa^2+\delta_1^2}-\frac{1}{\kappa^2+\delta_{-1}^2}} \label{eq:vtsideband}\\
&= s_0 \gamma_1 \sigma_1 J_0 \parlr{\frac{u_1}{\omega_B}} J_1 \parlr{\frac{u_1}{\omega_B}}8\kappa \omega_B \frac{\delta_0}{(\kappa^2+\delta_1^2)(\kappa^2+\delta_{-1}^2)} \label{eq:vtdel0sign}
\end{align}

From Equation (\ref{eq:vtsideband}), it is clear that the sign and magnitude of the transport velocity are set by the difference in the strength of the two sidebands at $\pm \omega_B$. In addition, Equation (\ref{eq:vtdel0sign}) clearly demonstrates that the direction of the transport can be tuned by choosing the sign of the effective detuning $\delta_0$. \mbox{In Figure \ref{fig:fig5}}, we give a quantitative comparison of the transport velocity obtained from the analytical calculation above with that obtained by a numerical calculation and find very good agreement. \mbox{We note} that our treatment above can be extended by including more than nearest neighbour couplings for the WS states (for example, in the relation determining $\mathcal{C}$ in Equation (\ref{eq:cos2xWS})), allowing us to describe the effects of the sidebands at higher multiples of $\omega_B$. We do not carry out this extension, as even at this level of approximation, the analytical theory agrees rather well with the numerical simulations.

Finally, it is clear from Equations (\ref{eq:centroidfinal}) and (\ref{eq:vtsideband}) that both the oscillatory motion and coherent transport are absent when the magnitude of the site-to-site coherence of the initial wave function $\sigma_1$ goes to zero. This serves to explain the behaviour observed in Figures \ref{fig:fig2b} and \ref{fig:fig3} (the black curve), where the initial site-to-site coherence is highly suppressed, giving raise to only a very small modulation and no transport.

\begin{figure}[h]
\centering
\includegraphics[width=8cm]{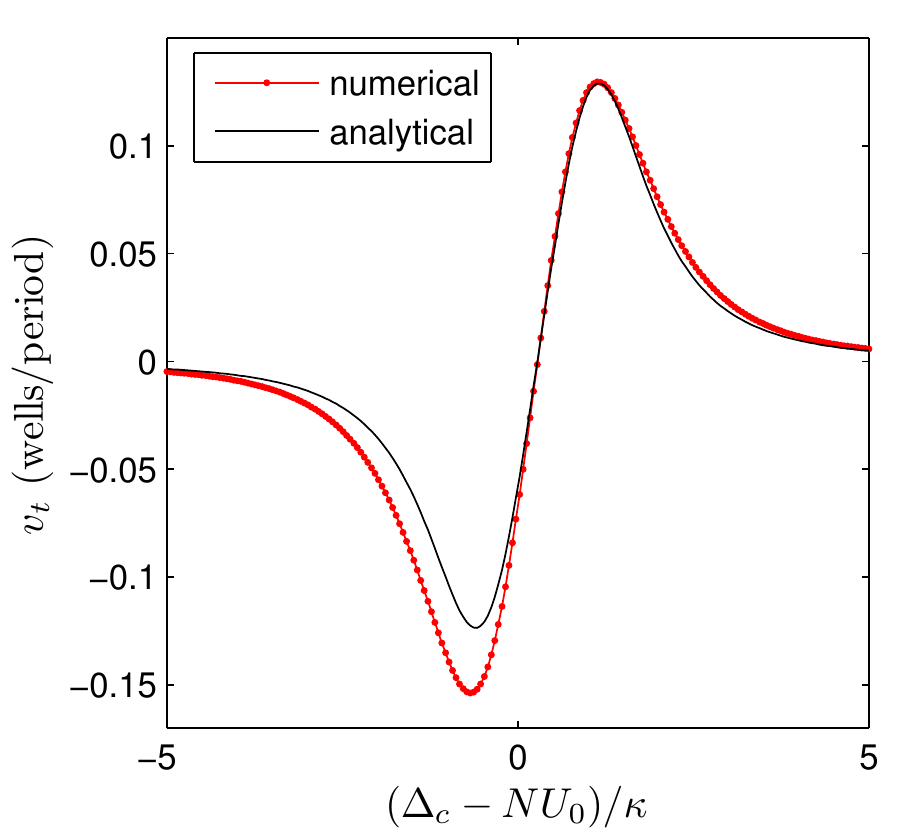}
\caption{Comparison of the transport velocity as a function of the cavity-driving laser detuning calculated from the numerical simulation and using the analytical expression Equation (\ref{eq:vtdel0sign}). System parameters are as introduced in Figure \ref{fig:fig2a} with $\eta$ varied to maintain an initial lattice depth of $-3 E_r$.}
\label{fig:fig5}
\end{figure}

\section{Metrology} \label{sec:metrology}

A fundamental property of Bloch oscillations is that their frequency depends only on the product of the applied force and the spatial periodicity of the optical lattice and not on the lattice depth. As we showed in Section \ref{sec:standard}, this remains true even when the nonlinearity due to dynamical backaction is included and the equations of motion must be solved self-consistently; the optical spring effect whereby the frequency depends on the amplitude of the cavity field is absent for the Bloch oscillator. This is good news for metrological applications and implies that continuous measurements of forces via Bloch oscillations of atoms in a cavity should be just as robust as their free-space (destructive) counterparts \cite{Ferrari06,Alb10,Poli11,Tar12}.

For metrology, it is necessary to go one step further and ask if there are any circumstances under which the Bloch frequency might be susceptible to a systematic shift. The most obvious way for this to happen is if the wavelength of the light in the cavity were to be different from that of the pump laser (which is assumed to be accurately known). A quick consideration of our theoretical model for the atom-cavity system, as described in Section \ref{sec:sec2}, reveals that it does not take into account any change in the spatial mode structure of the cavity field due to its interaction with the atoms. \mbox{The model }we use is the standard one adopted in theoretical treatments of dilute gases of atoms in optical \mbox{cavities \cite{Ritsch13}} and accurately describes the frequency shifts due to a dispersive interaction with atoms, but it does not include wavelength changes, because these are typically very small. The underlying physics is the phenomenon of resonance: we are typically only interested in frequencies in the vicinity of a cavity resonance, because that is where a significant amount of light is allowed into the cavity. In a good cavity, the width $\delta \lambda$ of a resonance is narrow in comparison to the natural scale over which the wavelength changes, which is given by the free spectral range $\Delta \lambda$. More precisely, the free spectral range is the wavelength separation between neighbouring resonances (which differ by one half wavelength fitting into the cavity), and its ratio to the resonance width is the finesse $\mathcal{F}=\Delta \lambda / \delta \lambda$, which can be a large number. For example, the cavity used in the experiment \cite{Brennecke08} was $178 \, \mu$m long with a power decay rate of $2 \pi \times 1.3$ MHz, giving $\mathcal{F}=$ 324,000. Thus, even if the dispersive interaction with the atoms shifts the resonance by many cavity linewidths, the effect on the wavelength will still be tiny. Conversely, the cavity resonances are super-sensitive to changes in the dispersive coupling. In terms of the refractive index $n_{r}$, the wavelength of light in a refractive medium shrinks according to $\lambda = \lambda_{0}/n_{r}$, where $\lambda_{0}$ is the vacuum wavelength, so that the change in refractive index required to shift the wavelength by one linewidth is \scalebox{.95}[1.0]{$\delta n_{r}= (d n_{r} / d \lambda) \, \delta \lambda = (d n_{r} / d \lambda) \, (d \lambda/ d \omega_{L}) 2 \kappa = (2/n_{r}) \kappa/ \omega_{L} \approx 2 \kappa/ \omega_{L} \approx 10^{-9}$,} where we have assumed that the pump laser frequency $\omega_{L}$ is in the THz range.

For completeness, let us estimate the refractive index due to the atoms in the cavity. In a dilute gas, the refractive index is related to real part $\chi'$ of the susceptibility as $n_{r} \approx 1 + \chi' / 2$. For a two-level atom \cite{cctbook}:
\begin{equation}
\chi'=- \frac{N}{V}\frac{d_{a}^2}{\epsilon_{0} \hbar} \frac{\Delta_{a}}{\Delta_{a}^2 + \gamma^2/4 + \Omega_{1}^2/2 } \approx
- \frac{N}{V}\frac{d^2}{\epsilon_{0} \hbar} \frac{1}{\Delta_{a}}
\end{equation}
where $\Omega_{1}=\Omega_{0} \langle \hbd \hb \rangle $ is the Rabi frequency and $N/V$ is the atom density. The second (approximate) equality holds in the large detuning regime. Comparing this expression with that for the optical lattice potential $U_{0}$ given in Section \ref{sec:sec2}, we see that we can write:
\begin{equation}
\chi'=-\frac{V_{c} }{V} \frac{ N \, U_{0}}{\omega_{c}} \
\end{equation}

If we take $N$ to be the number of atoms inside the mode volume, then $V_c/V=1$. Assuming this to be the case and putting $-N U_{0} = 2 \pi \times 1$ kHz, like in the simulations shown in Figure \ref{fig:fig2}, give a correction to the vacuum value $n_{r}=1$ of order $10^{-9}$. Although these estimates are rather crude in nature and assume, for example, that the atoms are homogeneously distributed, they all point to the same conclusion that the shift in the wavelength is negligible, even for the purposes of metrology.

{The precision of a frequency measurement improves as the observed signal is integrated for longer times, but only up to the coherence time, beyond which no further benefit is derived \cite{ClerkRMP2010}. An estimate of the coherence time for atomic motion in a cavity can be obtained by considering the momentum diffusion rate of a single atom belonging to an ensemble of atoms \cite{Hechen98,Fischer01}. Although this calculation treats the atomic centre of mass degrees of freedom as classical variables, it is interesting to note that estimates based on it agree well with heating rate measurements in the experiment \cite{Colombe07}. In the limit of large detuning $\Delta_a$ (low saturation), the momentum diffusion coefficient is \mbox{given by}:
\begin{align}
D_{\mathrm{tot}} \approx D_{\mathrm{sw}} \left[ 1 + 2 C \sin^2(2kz) \frac{\kappa^2}{\kappa^2+\Delta_f^2}\right] \label{eq:diffrate}
\end{align}
where $D_{sw}$ is the spontaneous emission rate at an antinode of the intracavity lattice given by \scalebox{.95}[1.0]{$D_{sw} = \hbar^2 k^2/2 \tau_{\mathrm{sp}}$,} with $\tau_{\mathrm{sp}}^{-1} = 2 \gamma \vert \alpha \vert^2 \Omega_{0}^2/\Delta_{a}^2$. Defining the coherence time $\tau$ as the time when the momentum distribution has a width equal to one half of the first Brillouin zone \emph{i.e.}, $k$, we find $\tau = \tau_{\mathrm{sp}}/[1 + 2 C \sin^2(2kz) \frac{\kappa^2}{\kappa^2+\Delta_f^2}]$. Note that replacing $\sin^2(2kz)$ by $1/2$ and $\Delta_f$ by zero reproduces the expression given earlier in the Introduction and also in \cite{Ven09}. For our estimate here, we replace the classical variables in Equation (\ref{eq:diffrate}) by their time-averaged mean-field values (since we are always in a time-dependent situation due to the Bloch oscillation dynamics), \emph{i.e.}, \scalebox{.95}[1.0]{$\sin^2 (2kz) \rightarrow \overline{\langle \sin^2 (2 kz) \rangle} \approx 0.5$ and} $\Delta_f \rightarrow \overline{\Delta_f}$. For the parameters used in Figure \ref{fig:fig2} with $1000$ atoms, $U_0 = -2\pi \times 1$ Hz, cooperativity $C = 1.3$, linewidth of $\gamma = 2\pi \times 7.6$ kHz corresponding to the $^1S_0 - ^3P_1$ transition in $^{88}$Sr and assuming a laser-atom detuning $\Delta_a = - 2\pi \times 10$ MHz, we find a coherence time of approximately $5$ ms. This is not very large compared to the Bloch oscillation period. Fortunately, as pointed out in \cite{Ven09}, the key parameters only appear in a certain combination, which can be used to rescale them without changing the dynamics and, yet, improve the situation significantly. In other words, the mean-field calculations performed in this paper are unchanged under a simultaneous scaling of $\Delta_a$ by a positive factor $r$ (which scales $U_0$ by $1/r$), of the number of atoms $N_a$ by the same factor of $r$ and of the driving strength $\eta$ by $\sqrt{r}$. This scaling maintains the same average intracavity lattice depth, but even a modest value of $r=20$, which gives a detuning of $\Delta_a = -2\pi \times 200$ MHz and 20,000 atoms, leads to the much larger coherence time of $2$ s (corresponding to thousands of Bloch oscillations) and is comparable to the estimate in \cite{Ven09}. Thus, in principle, a high-precision measurement of the Bloch frequency can be made by integrating the signal over such time scales.}

Let us close with some remarks on fundamental aspects of metrology using a Bloch oscillator that are suggested by the particular form of the optomechanical Hamiltonian given in Equation (\ref{eq:blochoscillator}). When $\delta_{0}=0$, the only term in the Hamiltonian depending on the light is the interaction term:
\begin{equation}
\hat{H}_{\mathrm{int}}=U_0 \gamma_1 \vert \alpha_{0} \vert (\hb+\hbd) (\hb_M + \hbd_M)
\end{equation}
which implies that the ``position'' of the Bloch oscillator $\hat{z} \propto (\hb_M + \hbd_M)$ leads to a phase shift of the cavity field \cite{Asp14}. Thus, rather than measuring the intensity modulation of the light in order to monitor the Bloch oscillation frequency, when $\delta_{0}=0$, one should instead perform homodyne detection and measure the phase modulation. Indeed, at zero (time averaged) total detuning from cavity resonance, the modulation of the detuning about the resonance peak because of the Bloch oscillations has no effect on the cavity field intensity to first order (assuming a symmetric line shape). This means that provided we work in the weak coupling regime where the modulation is small, there is vanishing backaction on the atoms, because they are only sensitive to light intensity and not phase. This would eliminate deleterious effects, such as parametric heating of the atoms due to being in a lattice with a time-modulated amplitude. However, even if the classical backaction can be eliminated in this way, this still leaves
quantum measurement backaction. In the quintessential case of the continuous measurement of the cavity mirror position by homodyne detection of the optical phase shift, quantum measurement backaction takes the form of an increased noise in the mirror momentum, in agreement with Heisenberg's uncertainty relation \cite{Asp14}. This feeds back into the mirror position and enforces the standard quantum limit, such that the measurement adds at least zero-point position noise to the intrinsic noise of the oscillator. However, there is an interesting twist in the case of the Bloch oscillator, because the ``position'' operator $\hat{z} \propto (\hb_M + \hbd_M)$ and ``momentum'' operator $\hat{p} \propto i(\hbd_M - \hb_M)$ commute according to Equation (\ref{eq:bocomm2}). This suggests that the Bloch oscillator quadratures are not subject to the usual Heisenberg uncertainty relations, and so, there could potentially be no quantum measurement backaction in this type of measurement. Whether or not this tentative conclusion holds up to further scrutiny would be worthy of further investigation.

\section{Conclusions} \label{sec:conclusions}

In this paper, we have developed an optomechanical description of an atomic Bloch oscillator in a cavity and compared and contrasted it with a standard optomechanical system consisting of a cavity with a mobile end mirror suspended on a spring. The quantum Hamiltonians for the two systems can be put into the same form, and the energy levels of both types of oscillator form ladders with equal spacing between excitations. However, the Wannier--Stark ladder for the Bloch oscillation case has no ground state (for an infinite lattice), whereas the harmonic oscillator ladder for the mirror-on-a-spring case is semi-infinite and has a ground state. Both systems can be cooled or heated in the resolved sideband regime by detuning the pump laser to near the appropriate sideband. In the case of the Bloch oscillator, cooling/heating corresponds to net transport of the atoms down/up the tilted lattice, much like an elevator, and this is achieved by red or blue detuning, respectively. We have found a fully-analytical expression for the transport velocity that agrees well with numerical simulation.

The potential energy lost/gained by the atoms during transport is extracted from the light, and this leads to asymmetric sidebands, as power is taken from higher sidebands and moved to lower ones in the case of uphill transport and \textit{vice versa} for downhill transport. This mechanism is further confirmed by force \textit{versus} displacement plots, where the dynamics trace out loops whose enclosed area gives the work done on the atoms by the light. We also find that initial conditions play a significant role in determining the dynamics: in the case of an initial atomic wave packet extending over many sites, we can obtain transport, but when the initial wave packet is localised to a single site, we find that the wave packet undergoes a breathing motion with no transport. The transport and breathing dynamics we find can all be obtained in free-space optical lattices by imposing amplitude or frequency modulation from the outside \cite{Alb10,Poli11,Tar12}, but in the cavity case, they are self-induced by dynamical backaction.

There are some significant differences between the Bloch oscillator and the harmonic oscillator. {Chief among these is that the backaction does not alter the frequency of the Bloch oscillations. \mbox{By contrast,} in the harmonic oscillator case, there is the so-called optical spring effect, which gives a dependence of oscillator frequency on field amplitude and detuning. To be clear, other motional frequencies are altered: because the intracavity lattice depth is modulated by the backaction, this will affect certain types of atomic motions, such as the oscillation frequency of an atom about the bottom of one of the potential minima \cite{Horak00}. Nevertheless, the Bloch oscillation frequency is robust against this depth modulation, because it only depends on the lattice period, not its depth.} It can be shifted if the wavelength of the light in the cavity is altered by its interaction with the atoms, but this effect is generally so tiny as to be completely negligible even by metrological standards. This is promising for applications where the optomechanical Bloch oscillator is used for continuous measurements of forces. Furthermore, the unusual commutation relations obeyed by the annihilation and creation operators for the Bloch oscillator suggest that homodyne measurements of the Bloch frequency might even evade quantum measurement backaction.


\vspace{6pt} 

\acknowledgments{\textbf{Acknowledgements:} Duncan H.J. O'Dell acknowledges the Natural Sciences and Engineering Research Council of Canada (NSERC) for funding. Jonathan Goldwin acknowledges funding from the Engineering and Physical Sciences Research Council of the U.K. (EPSRC) 
 (EP/J016985/1).}


\authorcontributions{\textbf{Author Contributions:} All three co-authors conceived of the idea for this paper together and also jointly interpreted the results. B. Prasanna Venkatesh provided the majority of the analytic calculations with some contributions provided by \mbox{Duncan H.J. O'Dell} and Jonathan Goldwin. B. Prasanna Venkatesh and Jonathan Goldwin performed the numerical simulations. The manuscript was primarily written by B. Prasanna Venkatesh with editing by all co-authors.}


\conflictofinterests{\textbf{Conflicts of Interest:} The authors declare no conflict of interest. }

\bibliographystyle{mdpi}
\renewcommand\bibname{References}

}
\end{document}